\documentclass[conference]{IEEEtran} 

\usepackage{fancyhdr, epsfig, epsf, amsthm, amsmath, amssymb, amsfonts, subfigure, color}
\usepackage{dblfloatfix} 
\usepackage{threeparttable}
\usepackage[noadjust]{cite}
\usepackage{dsfont}
\usepackage{enumerate}
\usepackage{comment}
\newtheorem{theorem}{Theorem}

\newtheorem{lemma}{Lemma}
\newtheorem{corollary}{Corollary}

\newtheorem{proposition}{Proposition}

\def\qed{$\blacksquare$}
\def\proof{\noindent{\emph{Proof:} }}

\def\endproof{\hfill \qed}

\def\E{\mathsf{E}}
\def\SINR{\mathsf{SINR}}

\def\l{\left}
\def\r{\right}
\def\({\left(}
\def\){\right)}

\def\[{\left[}
\def\]{\right]}

\newcommand{\nn}{\nonumber}

\def\EE{\mathsf{EE}}

\def\papertitle{Spatio-Temporal Network Dynamics Framework for Energy-Efficient Ultra-Dense Cellular Networks}
\IEEEoverridecommandlockouts

\begin{document}

\title{ \fontsize{19.6}{21}\selectfont \papertitle}

%
%
%

\author{Jihong Park, Seong-Lyun Kim, Mehdi Bennis$^{\dagger}$, and M\'erouane Debbah$^{\circ}$
\thanks{J. Park and S.-L. Kim are with School of Electrical \& Electronic Engineering, Yonsei University, Seoul, Korea (email: \{jhpark.james, slkim\}@ramo.yonsei.ac.kr). M. Bennis$^{\dagger}$ is with Centre for Wireless Communications, University of Oulu, Finland (email: bennis@ee.oulu.fi). M. Debbah$^{\circ}$ is with Mathematical and Algorithmic Sciences Lab, France Research Center, Huawei Technologies Co. Ltd. (email: merouane.debbah@huawei.fr).}}

\maketitle \thispagestyle{empty}

\begin{abstract}
This article investigates the performance of an ultra-dense  network (UDN) from an  energy-efficiency (EE) standpoint leveraging the interplay between stochastic geometry (SG) and mean-field game (MFG) theory. In this setting, base stations (BSs) (resp. users) are uniformly distributed over a two-dimensional plane as two independent homogeneous Poisson point processes (PPPs), where users associate to  their nearest BSs. The goal of every BS is to maximize its own energy efficiency subject to channel uncertainty, random BS location, and interference levels. Due to the coupling in interference, the problem is solved in the mean-field (MF) regime where each BS interacts with the whole BS population via time-varying MF interference. As a main contribution, the asymptotic convergence of MF interference to zero is rigorously proved in a UDN with multiple transmit antennas. It allows us to derive a closed-form EE representation, yielding a tractable EE optimal power control policy. This proposed power control achieves more than 1.5 times higher EE compared to a fixed power baseline.
\end{abstract}


\section{introduction}
The relentless surge in traffic demands  has compelled network operators to seek innovative solutions to maximize their performance in terms of users' data rates and energy expenditures \cite{Andrews5G:14,FiveDisruptive:14}. In parallel to that, $5$G systems are expected to be ultra-dense in nature, i.e. $\lambda_b \gg \lambda_u$ \cite{CLI:16,Youssef:16,JHParkTWC:15}, rendering network optimization highly complex \cite{Andrews5G:14, Alexiou13}. Resource management techniques such as power control and scheduling in ultra-dense networks (UDNs) become significantly more challenging than in traditional sparse deployments due to spatio-temporal traffic demand fluctuations and channel uncertainties. Current state-of-the-art approaches are merely based on heuristics or system-level simulations lacking fundamental insights.  

Recently, mean-field games (MFGs) received significant attention in cellular networks based on the fact that players make their  local decisions (based on their own states) while abstracting other players' strategies using a mean-field (MF) measure obtained after solving a set of coupled partial differential equations (PDEs), known as Hamilton-Jacobi-Bellman (HJB) and Fokker-Planck-Kolmogorov (FPK) \cite{Meriaux:13,Tembine:11, MBennisGC:15}. Nonetheless, solving a set of coupled PDEs is still cumbersome, so none of the current MFG works incorporate spatial aspects of practical networks such as the randomness of BS locations.

From a stochastic geometry (SG) perspective, while there is a significant body of literature in terms of performance modeling and analysis, most preceding works overlook the temporal network dynamics. Such missing gaps motivate this work which aims at capturing the spatio-temporal nature of wireless networks. To the best of our knowledge, this is the first work combining MFG and SG frameworks within the scope of UDNs.

Our contributions are as follows. Firstly, MF Interference convergence is rigorously proved (Theorem 1), i.e. interference normalized by BS density is finite in UDNs. The result specifies necessary conditions: (i) BS densification (scaling law), (ii) user density, (iii) fading channel, and (iv) LOS dependency, captured by a so-called reception ball size. Secondly, an optimal power control policy that maximizes EE is obtained (Proposition 2) via a closed-form expression of energy-efficiency (EE) (Proposition 1) by leveraging the MF measure and MF interference convergence.


\section{System Model} \label{Sect:SysModel}
Consider a downlink cellular network where base stations (BSs) are uniformly distributed over a two-dimensional infinite Euclidean plane with density $\lambda_b$, leading to a homogeneous Poisson point process (PPP). Similarly, users are uniformly distributed with density $\lambda_u$, independent of BS locations. Each user associates with the \emph{nearest} BS. This results in a set of geographical coverages of BSs, i.e. Voronoi tessellation \cite{StoyanBook:StochasticGeometry:1995}. A BS having no serving user within its coverage becomes \emph{dormant}, not transmitting any signals so as to save energy as well as to mitigate interference; otherwise, \emph{active}.

A single active BS directionally transmits a data signal to the associated user by using $N$ number of antennas. Its beam pattern experienced at a receiving user follows a sectorized uniform linear array antenna model \cite{HeathWearable:15} where the main lobe gain $G_N$ is $N$ with beam width $\theta_N=2\pi/\sqrt{N}$, neglecting side lobes and assuming beam centers point at the associated users. 

Transmitted signals from BSs experience path-loss attenuation. The attenuation from the $i$-th BS coordinates $z_i$ to the $j$-th user coordinates $y_j$ $l_{ij}{(\lambda_b)}=\min\(1, \|z_i-y_j\|^{-\alpha}\)$ where $\alpha>2$ denotes the path-loss exponent. Its bounded unity value comes from the fact that path-loss attenuation cannot amplify the transmitted signal. In addition to path-loss attenuation, the transmitted signals at time $t\geq 0$ experience spatially i.i.d. but temporally correlated fading with the fading coefficient vector $g_{ij}(t)=\{g_{x_{ij}}(t), g_{y_{ij}}(t)\}$. This fading gain is constant within a block, and evolves from block to block while satisfying the Markov property as in a Gauss-Markov correlated block fading model \cite{Honig:12}. Consider the block length is infinitesimal, leading to the following fading evolution law \cite{Meriaux:13}:

\vspace{-5pt}\small\begin{align}
dg_{ij}(t)&= \frac{1}{2}\(\mu - g_{ij}(t)\) dt + \eta d\mathbb{W}_{ij}(t) \label{Eq:Channel}
\end{align}\normalsize
where $\mu:=\{\mu_x, \mu_y\}$ for non-negative constants $\mu_x, \mu_y$, $0\leq \eta < \infty$, and $\mathbb{W}_i(t)$'s are mutually independent Wiener processes. Adjusting $\mu$ and $\eta$ respectively determines the amounts of temporal correlation and (slow/fast) fading variance (see \cite{Meriaux:13} for further details). Combining the effects of path-loss attenuation and fading yields channel gain $h_{ij}$ from the $i$-th BS to the $j$-th user, given as $|h_{ij}(t, \lambda_b)|^2 = l_{ij}{(\lambda_b)} \times |g_{ij}(t)|^2$. In the following, we abuse notation by using the sole subscript $i$ for indicating the associated BS $i$ to user $i$ links.

The $i$-th BS transmission power $0 <P_i\l\{t, h_i(t, \lambda_b)\r\} \leq P_{\max}$, which is determined only by its own channel state in $h_i(t, \lambda_b)$. Note that different BSs having identical channel state transmit the same power, i.e. homogeneous admissible controls for BS transmission powers.

User $i$ only receives signals from BSs located within a \emph{reception ball} $b(y_i, R)$ centered at $y_i$ with radius $R>0$. Motivated by the line-of-sight (LOS) ball model \cite{Heath:13}, adjusting $R$ enables emulating not only sub 6 GHz cellular networks but also LOS dependent networks where guaranteeing LOS is necessary for communication. For LOS dependent networks such as above 6 GHz millimeter-wave cellular networks, $R$ corresponds to the average LOS distance that can be calculated at a given geographical site \cite{Heath:13,JHParkTWC:15}. For sub 6 GHz networks, $R$ corresponds to the average distance that provides he received signal power larger than noise floor. When $R\rightarrow \infty$, the reception ball model becomes identical to a traditional PPP network model \cite{Andrews:2011bg}.

\section{Mean-Field Interference in Ultra-Dense Cellular Networks}
Interactions of BSs through interference is a major source of the bottlenecks for optimizing wireless networks. MFG allows us to circumvent this problem by changing the entire interactions into a single interaction via MF interference. To achieve this, MF interference should asymptotically converge to zero, verified as in the following.

Let $\Phi_R(\lambda)$ denote either (i) point coordinates following homogeneous PPP with density $\lambda$ within $b(o, R)$ or (ii) the number of such points within $b(o, R)$. Consider a randomly selected user, i.e. typical user. Now that the user coordinates are translation-invariant, i.e. stationary, it is always possible to make this typical user located at the origin $o$, i.e. Slyvnyak's theorem \cite{StoyanBook:StochasticGeometry:1995}. At the typical user, the following statements hold according to the nearest association and user-void dormant BSs under the reception ball model in Section \ref{Sect:SysModel}.

\begin{itemize}
\item If at least a single BS exists within $b(o, R)$ with probability $p_R$, the associated BS is the nearest out of the entire BS coordinates $\Phi_R(\lambda_b)$ (recall the associated BS is always active).

\item An arbitrary interfering BS $i$ belongs to the set of active BS coordinates $\Phi_R(p_a\lambda_b)$ where $p_a$ denotes a BS's active probability.
\end{itemize}

Utilizing the void probability of a PPP \cite{StoyanBook:StochasticGeometry:1995} yields $p_R=1-\exp\(-\pi \lambda_b R^2\)$. Assuming $p_a$'s are homogeneous over BSs provides $p_a \approx 1-\[1+\lambda_u/(3.5\lambda_b)\]^{-3.5}$ \cite{Yu2011}. For sufficiently large $\lambda_b/\lambda_u$, the accuracy of this approximation is in \cite{SLeeKHuang12}, and asymptotic convergence is verified in \cite{JHParkTWC:15}.

Let the subscript $0$ denote the associated BS. At a typical user, the received power from the associated BS and aggregate interfering BSs with a single antenna are respectively given as follows:
\begin{align}
S_{\lambda_b}(t) &= P_0\l\{t, h_0(t, \lambda_b)\r\} |h_0(t, \lambda_b)|^2 \\
I_{p_a \lambda_b} (t) &=  \sum_{i= 1}^{|\Phi^o_R(p_a \lambda_b)|} P_{i}\l\{t, h_{i}(t, p_a\lambda_b)\r\}|h_{i0}(t, p_a\lambda_b)|^2
\end{align}
where $\Phi^o_R(p_a \lambda_b) := \Phi_R(p_a \lambda_b)\backslash\{z_0\}$ via Slyvnyak's theorem.

Signal-to-interference-plus-noise ratio ($\SINR$) with $N$ number of transmit antennas is then:

\begin{align}
\SINR(t) &= N \cdot S_{\lambda_b}(t) / \bigg( \sigma^2 + \frac{ \theta_N}{2\pi}N\cdot I_{p_a\lambda_b}(t)\bigg)\\
&= S_1(t)/ \bigg( \frac{\sigma^2}{N\sqrt{{\lambda_b}^\alpha}} + \underbrace{\frac{{I}_{p_a\lambda_b}(t)}{\sqrt{N{\lambda_b}^\alpha} }}_{\hat{I}_{p_a\lambda_b}(t)} \bigg) \quad a.s. \label{Eq:SINR}
\end{align}\normalsize
where $\sigma^2$ noise power and the last step follows from $S_{\lambda_b}(t) \sim {\lambda_b}^\frac{\alpha}{2} S_1(t)$ via mapping theorem \cite{HaenggiSG}. Note that $\hat{I}_{p_a\lambda_b}(t)$ is the interference \emph{normalized by BS density and the number of antennas}.

\subsection{MF Interference}
Our objective is to prove the normalized interference $\hat{I}_{p_a\lambda_b}(t)$ in \eqref{Eq:SINR} asymptotically converges to a finite value, i.e. MF interference. According to \cite{Meriaux:13}, it requires to satisfy the following three conditions.
\vspace{5pt}\begin{enumerate}[1.]
\item \textsf{Exchangeability} -- The power control of $P_i\{t, h_i(t, \lambda_b)\}$'s is invariant by permuting channel states $h_i$'s. Guaranteeing this condition allows us to consider a generic interfering BS with the channel state $x \in \chi$ where $\chi$ is the set of the entire states.
\item \textsf{MF measure existence} -- MF measure $m_t(x)$ exists for all $x \in \chi$, which is the channel state distribution function evolving over time.
\item \textsf{Finite MF interference} -- Normalized interference averaged over $x\in\chi$ via $m_t$ is bounded.
\end{enumerate}\vspace{5pt}

For a generic BS and its corresponding functions, the subscript $i$ is neglected henceforth. By verifying the three conditions (see Appendix for details), the following proposition provides MF interference while specifying its necessary conditions.

\begin{theorem} \emph{(MF Interference in UDNs)} In a downlink cellular network where its fading channel guarantees \emph{\textbf{A1}}. $\E_{g_i(t)} \[|g_i(t)|^2\] <\infty$, if either one of the following conditions is satisfied
\begin{description}
\item \hspace{-35pt}\emph{\textbf{A2}}. For $R<\infty$, $N {\lambda_b}^\alpha/{\lambda_u }^4 \rightarrow \infty$ and $\lambda_u \rightarrow \infty$
\item \hspace{-35pt}\emph{\textbf{A3}}. For $R\rightarrow\infty$, $N {\lambda_b}^\alpha/\( {\lambda_u R}\)^4 \rightarrow \infty$
\end{description}
then $\hat{I}_{p_a\lambda_b}(t)$ in \eqref{Eq:SINR} converges $a.s.$ to finite MF interference
\begin{align}
\hat{I}_1(t) &\sim  \E_\chi \[P\{t, x\} \cdot |h(t)|^2 \] \rightarrow 0 \label{Eq:MFinterf}
\end{align}
where $\E_\chi \[P\{t, x\} \cdot |h(t)|^2 \] = \int_{x\in\chi} P\{t, x\} |h(t)|^2 m_t(dx)$, $m_t=\lim\limits_{R\; or\; \lambda_u\rightarrow \infty}1/|\Phi_R^o(\lambda_u)|\sum\limits_{i=1}^{|\Phi_R^o(\lambda_u)|}\delta_{h_i(t, 1)}$ and $\delta_{h_i(t, \lambda_u)}$ Dirac measure concentrated at $h_i(t,\lambda_u)$.
\end{theorem}

The necessary conditions for MF interference convergence put emphasis on -- not BS density but -- the ratio of BS density to user density. As the ratio asymptotically goes to infinity MF interference converges. Such a condition interestingly corresponds with the definition of a UDN \cite{JHParkTWC:15}. Asymptotic increase in the number of antennas, such as massive MIMO, also ensures MF interference converges.

Unlike a large-scale network ($R\rightarrow \infty$), a finite network ($R<\infty$) as in millimeter-wave networks requires a huge number of users to guarantee MF interference convergence; otherwise, the number of active interferers become too small to formulate an MFG. This affects the necessary BS/user density ratio that requires more BS densification compared to a large-scale network.

Utilizing the step 3-1) in the proof of Theorem 1 yields the following closed-form MF interference.
\begin{corollary} \emph{(MF interference)} In a downlink cellular network, MF interference is given as follows:
\begin{itemize}
\item If \emph{\textbf{A1}} and \emph{\textbf{A2}} are satisfied under $R < \infty$

\vspace{-5pt}\small\begin{align}
\hspace{-10pt}\hat{I}_{p_a\lambda_b}(t) &=   \frac{(\lambda_u \pi R)^2}{\sqrt{N}{\lambda_b}^\frac{\alpha}{2}} \(1 + \frac{1-R^{2-\alpha}}{\alpha-2}\)\hat{P}(t)  \E_g\[|g(t)|^2\] ;
\end{align}\normalsize
\item If \emph{\textbf{A1}} and \emph{\textbf{A3}} are satisfied under $R  \rightarrow \infty$

\vspace{-5pt}\small\begin{align}
\hspace{-10pt}\hat{I}_{p_a\lambda_b}(t) &=   \frac{(\lambda_u \pi R)^2}{\sqrt{N}{\lambda_b}^\frac{\alpha}{2}} \(1 + \frac{1}{\alpha-2}\)\hat{P}(t)  \E_g\[|g(t)|^2\] 
\end{align}\normalsize
\end{itemize}
where $\hat{P}(t) = \int_{x\in\chi} P\{t, x\} m_t(dx)$.
\end{corollary}

It is worth noting that asymptotically vanishing MF interference in (6) indicates -- not interference but -- the normalized interference defined in (5) becomes zero. For the same reason, a UDN where MF interference converges can not only be noise-limited but also be interference-limited. In (5), as in interference, noise power is normalized also by the same value, i.e. BS density and the number of antennas. Determining either interference or noise-limited therefore depends on whether interference dominates noise power or nor, independent of the normalization.

To further facilitate the applicability of MF interference in a practical scenario, consider relaxing \textbf{A2} and \textbf{A3} in Theorem 1. In a downlink UDN with the relaxed non-asymptotic condition \textbf{A0}. $N {\lambda_b}^\alpha /(\lambda_u R)^4 \gg 1$, Fig. 1 depicts its average rate per unit spectrum bandwidth $\E\[1 + \SINR(t)\]$. It shows that both MF interference applied (solid blue) and simulated (dotted red) average rates increase along with the BS/user density ratio. This verifies the MF approximation achieves more than $98.3$\% accuracy, the ratio of the simulated to the MF approximated values. The MF approximation becomes tighter as the BS/user density ratio grows. On the basis of this validation, Section IV utilizes MF interference approximation for maximizing EE, which is linearly proportional to the average rate.

\begin{figure}\centering
\hspace{-10pt}\includegraphics[width=9cm]{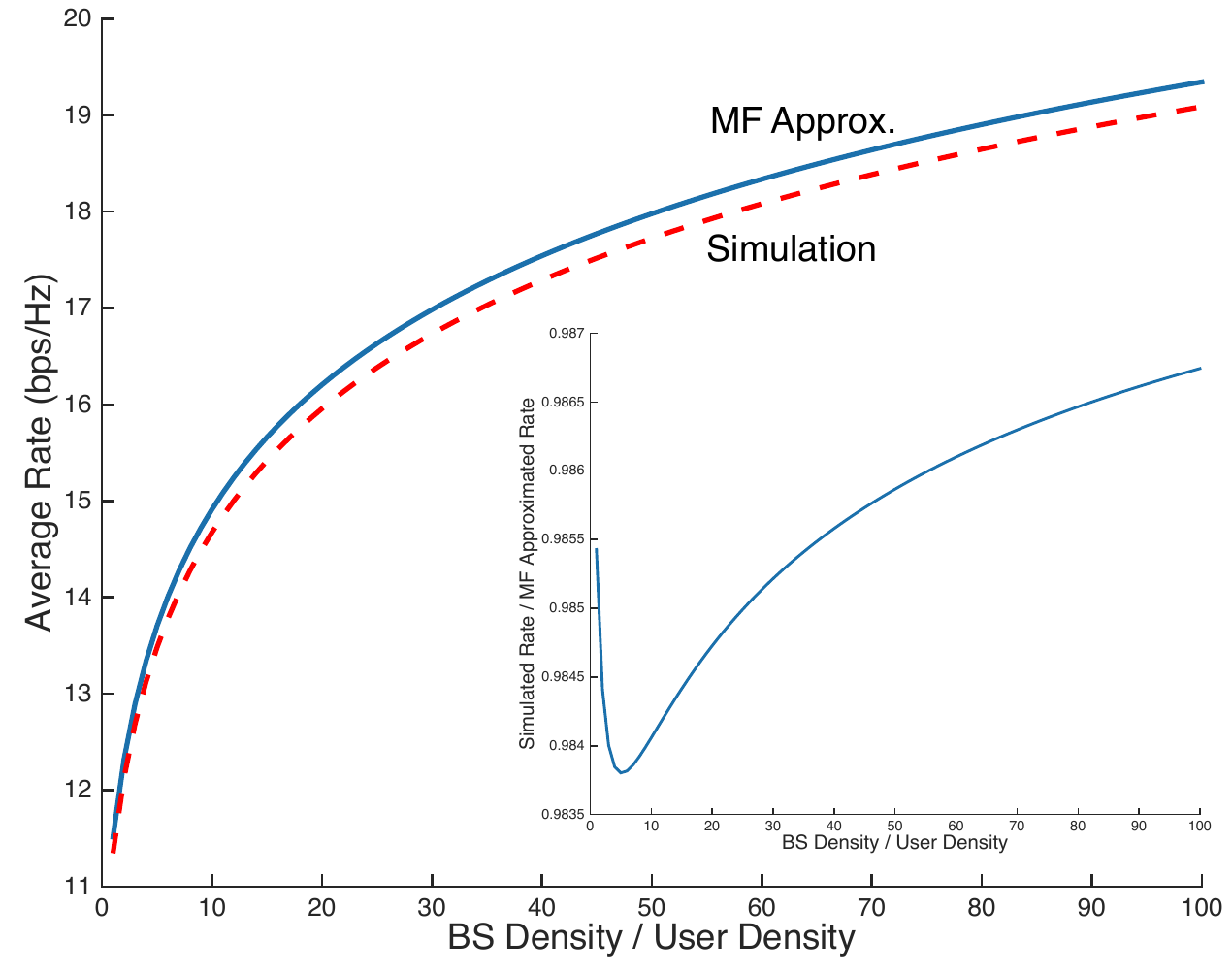}
\caption{Average rate with and without MF interference approximation ($\lambda_u=0.001$, $N=10$, $\alpha=4$, $\eta=1$, $|\mu|=\sqrt{2}$, $R\rightarrow \infty$, $\sigma^2=0.001$, $P_0(t)=P_i(t)=1$).}
\end{figure}

\subsection{MF Interference from the SG Point of View}
The proof of Theorem 1 in Appendix is based on a MFG formulation and its necessary conditions for the MF convergence. Instead, the following different proof sketch of Theorem 1 provides an interpretation from an SG standpoint.

At a typical user, applying \eqref{Eq:SINR} yields the average rate 

\vspace{-5pt}\small\begin{align}
\hspace{-10pt} \E_{S, I}\[ \log\(1+ \SINR(t)\)\] &=   \E_{S}\E_I \[\log \(1 + \frac{S_1(t)}{\frac{\sigma^2}{N\sqrt{{\lambda_b}^\alpha}}  + \hat{I}_{p_a\lambda_b}(t)} \)\] \label{Eq:II1}\\
&\hspace{-80pt}\leq   \frac{(m + M)^2}{4m M}\E_S \log \(1 + \E_I\[ \frac{S_1(t)}{\frac{\sigma^2}{N\sqrt{{\lambda_b}^\alpha}}  + \hat{I}_{p_a\lambda_b}(t)} \]  \) .  \label{Eq:II2}
\end{align}\normalsize

The first step follows from \textbf{A0}. In this case, such a decomposition is allowed since the nearest association of $S$ does not affect $I$, i.e. no interferer within the association distance from the typical user.

The last inequality comes from Kantorovich's inequality \cite{Kantorovich:48} for the convex function of $\hat{I}_{p_a\lambda_b}(t)$ where $m$ and $M$ respectively denote the minimum and maximum values of $\hat{I}_{p_a\lambda_b}(t)$. Its equality holds when $m=M$. According to \textbf{A0} and the definition of $\hat{I}_{p_a\lambda_b}(t)$, $m\approx M$, leading to approximating \eqref{Eq:II2} to \eqref{Eq:II1}.

Consider the term inside the logarithm in \eqref{Eq:II2}.

\vspace{-5pt}\small\begin{align}
\E_I\[ \frac{S_1(t)}{\frac{\sigma^2}{N\sqrt{{\lambda_b}^\alpha}} + \hat{I}_{p_a\lambda_b}(t)} \]
&\geq  S_1(t) \( \frac{\sigma^2}{N\sqrt{{\lambda_b}^\alpha}} - \E_I \[\hat{I}_{p_a\lambda_b}(t)\] \)    \label{Eq:II3}
\end{align}\normalsize
The above inequality follows from Taylor's expansion for $\hat{I}_{p_a\lambda_b}(t)$. Combining \eqref{Eq:II1}--\eqref{Eq:II3} consequently yields the result.

\vspace{-5pt}\small\begin{align}
\E_{S, I}\[ \log\(1+ \SINR(t)\)\] &\approx \E_{S}\log \(1 + \frac{S_1(t)}{\frac{\sigma^2}{N\sqrt{{\lambda_b}^\alpha}} + \E_I\[\hat{I}_{p_a\lambda_b}(t)\] } \) \label{Eq:II}
\end{align}\normalsize
Note that the approximation above becomes the exact calculation when applying \textbf{A2} and \textbf{A3} instead of \textbf{A0}.

The result \eqref{Eq:II} derived from a SG perspective interestingly corresponds to the Theorem 1 derived from a MFG point of view. The reason is taking expectation over states (i.e. MF) is identical to the spatial average with the marks other than the transmitter locations out of the states.

\section{Energy-Efficiency Maximization under Mean-Field Interference}
The goal of this section is to validate the effectiveness of the proposed approach. To this end, we maximize the energy efficiency of a UDN in a tractable manner by virtue of MF interference. For simplicity, this section only considers channel states, incorporating the impact of energy states is deferred to future work. In the following, we may abuse notations by dropping $i$, $t$ and/or $\lambda$. 

Consider at first an instantaneous energy efficiency $\EE(t)$ defined as the spatially averaged downlink capacity divided by power consumption at time $t$, i.e. $\EE(t):=\E_{y_i}[\log\(1+\SINR_i(t)\)]/(P_i(t)+P_c)$ where $P_c$ denotes the fixed circuit power consumption at BS $i$. In a UDN where MF interference approximation holds, $\EE(t)$ can be represented in closed-form as the following proposition shows.

\begin{proposition} \emph{(EE under MF interference)} In a downlink UDN where \emph{\textbf{A1}--\textbf{A3}} in Theorem 1 hold, $\EE(t)$ at a typical user at time $t$, is given as follows:

\vspace{-10pt}
\small\begin{align}
\EE(t) &= \frac{1}{P_c + P(t)}\[c_1   + \log\(\frac{P(t) }{\frac{\sigma^2}{N\sqrt{{\lambda_b}^\alpha}} + \hat{I}_{p_a\lambda_b}(t)} \)\]
\end{align}\normalsize  \label{Eq:Prop1}
where $c_1=2\E_{g}\[\log |g(t)| \] + \alpha\(\gamma + \log\pi\)/2 $ and $\gamma\approx 0.5771$ Euler constant.
\end{proposition}

\begin{figure*}
\setcounter{equation}{29}

\vspace{-5pt}\small\begin{align}
\frac{P_c}{P^*(t)} &=    W\(\frac{ P_c  e^{c_1-1}}{\frac{\sigma^2}{N\sqrt{{\lambda_b}^\alpha}} + \hat{P}^*(t) \[2 \eta^2\(1-e^{-t}\)^2 + |\mu|^2\(1-e^{-\frac{t}{2}}\)^2\]  \frac{(\lambda_u \pi R)^2}{\sqrt{N {\lambda_b}^\alpha}} \(1 + \frac{1}{\alpha-2}\)     } \) \label{Eq:Prop2}
\end{align}
where $W(y)$ is a Lambert W function providing the solution $x$ of $y=x e^x$, which monotonically increases for $y>0$ \normalsize

\hrulefill
\end{figure*}

\noindent\emph{Proof:} Regardless of being either noise-limited or interference-limited, $\SINR \gg 1$ holds in a UDN according to Theorem 1 and \eqref{Eq:SINR}, and thus $\log\(1+\SINR_i(t)\)\approx \log\(\SINR_i(t)\)$. Applying this to the counter CDF (CCDF) of energy efficiency at a typical user conditioned on the typical user's associated channel fading under a given MF interference $\hat{I}_1(t)$:

\setcounter{equation}{25}\vspace{-5pt}
\small\begin{align}
\hspace{-11pt}\Pr\(\frac{\log\(1+\SINR\)]}{P(t)+P_c} >v \mid g\) &= \Pr\(|z| < \[\frac{c_0 }{e^{v \(P(t)+P_c\)}}\]^{\frac{1}{\alpha}} \) \\
&\hspace{-26pt}= 1- \exp\(-\pi \[ c_0 e^{-v(P(t)+P_c)} \]^{\frac{2}{\alpha}}\)  \label{Eq:PfGumbel}
\end{align}\normalsize
where $c_0:= P(t) |g(t)|^2/ \(\frac{\sigma^2}{N\sqrt{{\lambda_b}^\alpha}} + \hat{I}\)$ and the last step follows from the void probability of the nearest BS \cite{StoyanBook:StochasticGeometry:1995}.

Notice \eqref{Eq:PfGumbel} is identical to the CCDF of a Gumbel distributed random variable with parameters $\beta = \[2(P(t)+P_c)/\alpha\]^{-1}$ and $\mu = \beta \log\(\pi {c_0}^{2/\alpha}\)$. Taking an integration over $v>0$ at \eqref{Eq:PfGumbel} thereby leads to the Gumbel mean given as $\mu + \beta \gamma$. Taking an expectation over the associated fading channel provides the spatially averaged EE by using Slyvnyak's theorem \cite{StoyanBook:StochasticGeometry:1995} \endproof

\vspace{5pt}
It is worth mentioning that replacing $P(t) + P_c$ by unity in Proposition 1 yields the instantaneous spatially averaged downlink rate. Respectively for noise-limited or interference-limited regime, this downlink rate is a logarithmic function of BS density or BS/user density ratio (recall the definition of $\hat{I}$), which is interestingly in line with our preceding result without incorporating MF interference \cite{JHParkTWC:15}.

Calculating $\EE(t)$ is viable by using $g(t)$'s evolutional law given as \eqref{Eq:Channel}. Given $t>0$, the following verifies $|g(t)|$ follows Rician fading.
\begin{lemma} \emph{(Temporal fading distribution)} When the initial condition $g(0)=0$ is satisfied, at $t>0$ Euclidean norm of the fading vector $|g(t)| \sim$ Rice$\(|\mu|\[1-e^{-\frac{t}{2}}\],\eta\[1-e^{-t}\]\)$.
\end{lemma}

\noindent\emph{Proof:} The evolutional law of $g(t)$ is an Ornstein-Uhlenbeck process whose probability density function (PDF) $f_g(u,t)$ satisfies the following FPK.
\begin{align}
\frac{df_g(u,t)}{dt} &= \frac{1}{2}\[(u-\mu)f_g(u, t)\] + \frac{\eta^2}{2}\frac{d^2 f_g(u,t)}{d u^2}
\end{align}

\noindent This provides $g(t)\sim \mathcal{N}\(\mu\[1-e^{-\frac{t}{2}}\], \eta^2 \[1-e^{-t}\]^2 \)$. The desired result is verified by exploiting the relationships: (i) $\sqrt{X^2 + Y^2} \sim \text{Rice}(\nu, \sigma)$ for $X \sim \mathcal{N}(\nu \cos\theta, \sigma^2)$ and $X \sim \mathcal{N}(\nu \sin\theta, \sigma^2)$ where $\theta\geq 0$ and (ii) $\sin^2\theta + \cos^2\theta=1$.
\endproof

Utilizing Proposition 1 with Lemma 1 allows us to consider the EE maximization at $t>0$.
\begin{align}
&\hspace{30pt}\underset{P(t)}{\textsf{maximize}}\quad  \EE(t)\\
&\textsf{s.t.}\quad dg(t)= \frac{1}{2}\(\mu - g(t)\) dt + \eta d\mathbb{W}(t) \nn
\end{align}

The solution of this problem is given in the following proposition.
\begin{proposition} \emph{(EE optimal power)} For $t>0$ at a typical user, the optimal power $P^*(t)$ maximizing $\EE(t)$ satisfies \eqref{Eq:Prop2} at the top of the next page.
\end{proposition}

\proof 
According to Proposition 1, $\EE(t)$ is an S-shaped unimodal function w.r.t. homogeneously controlled power $P(t)$, which has a unique solution. Applying the first-order necessary condition thereby yields the optimal power $P^*(t)$ as follows

\setcounter{equation}{30}\vspace{-12pt}\small\begin{align}
P^*(t) &= P_c\[W\(\frac{ P_c e^{c_1-1}}{\frac{\sigma^2}{N\sqrt{{\lambda_b}^\alpha}} + \hat{I}_{p_a\lambda_b}(t)  } \)\]^{-1}. \label{Eq:PfProp2}
\end{align}\normalsize

\noindent Note that $\hat{P}(t)$ in $\hat{I}_1(t)$, interpreted as a single MF interferer's transmission power, interacts with $P^*(t)$ in \eqref{Eq:PfProp2} under the homogeneous power control. This recursion converges to a unique $\hat{P}^*(t)$ within a couple of iterations (see Fig. 2). The proof is completed by applying Corollary 1 and Lemma 1.
\endproof

\begin{figure} \centering
\includegraphics[width=7.7cm]{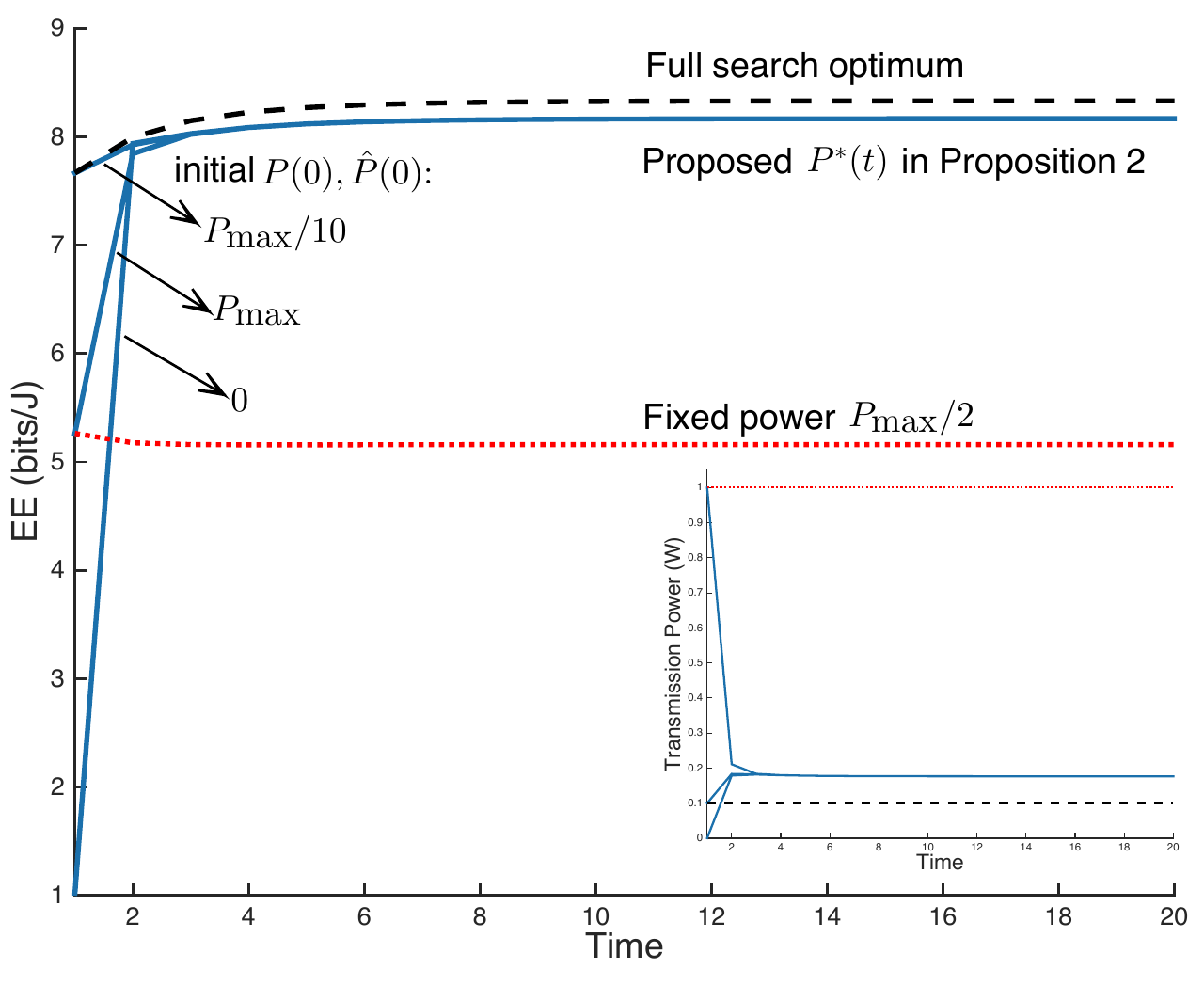}
\caption{Maximized EE by the proposed transmission power control in Proposition 2, compared with fixed power and full search optimum transmissions ($\lambda_b=10$, $\lambda_u=1$, $N=1$, $P_{\max}=1$, $P_c=1$, $\eta=1$, $|\mu|=\sqrt{2}$, $\alpha=4$).}
\end{figure}

\begin{figure}\centering
\includegraphics[width=8.5cm]{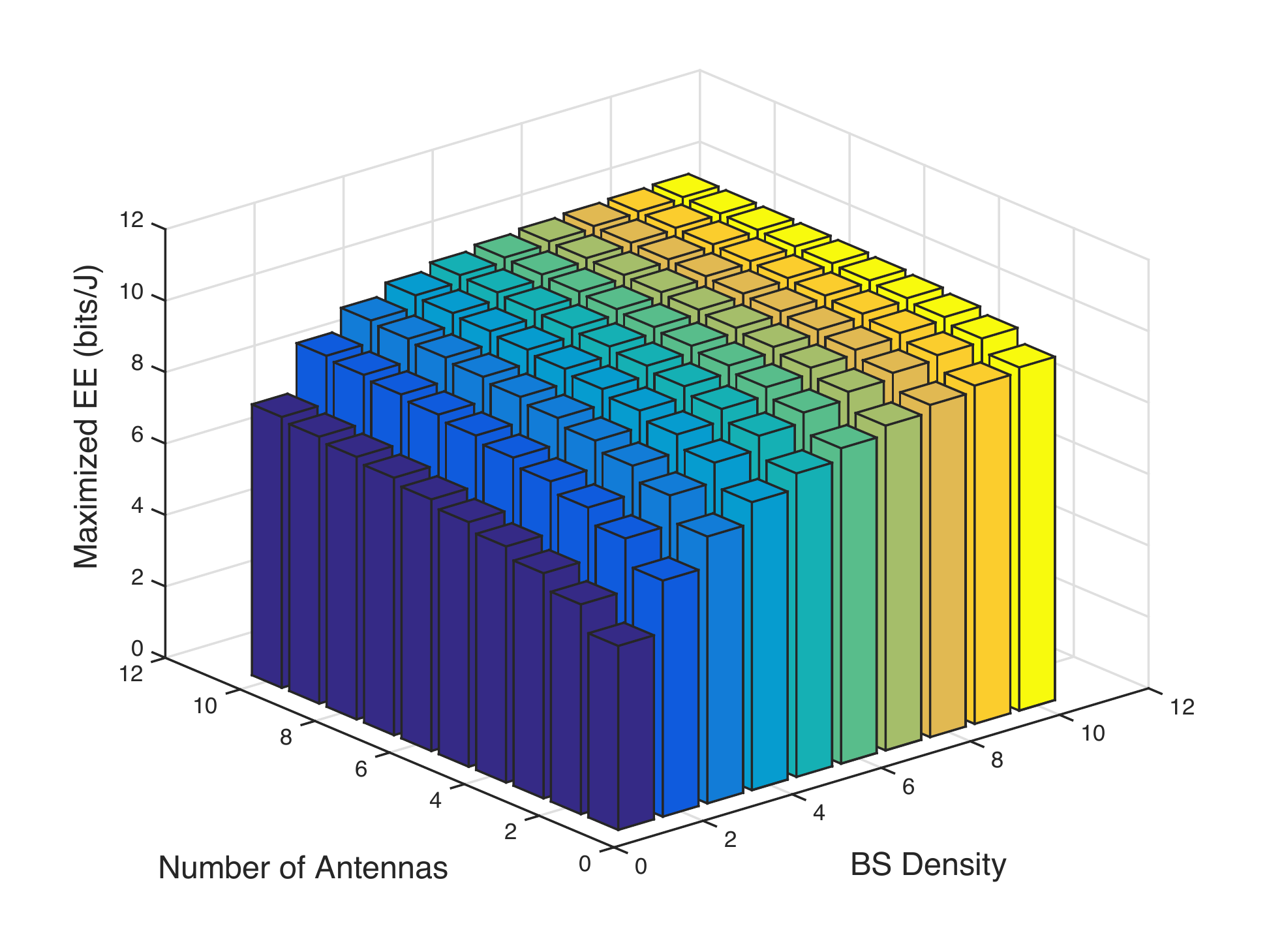}
\caption{Maximized stationary EE by the proposed transmission power control in Proposition 2 with respect to the number of antennas and BS density ($P_{\max}=1$, $P_c=1$, $\eta=\mu_x=\mu_y=1$, $\alpha=4$, $R=10$).}
\end{figure}

\section{Numerical Evaluation}
This section validates the impact of the proposed power control policy under MF interference (Proposition 2).

Fig. 2 shows the proposed power control (solid blue) provides $1.58$ times higher EE compared to a $P_{\max}/2$ fixed power baseline (dotted red), which corresponds to achieving $98$\% of the EE maximum via a full search algorithm (dashed black). For small $t$, transmission power of the proposed scheme fluctuates due to the recursive relationship between $P(t)$ and $\hat{P}(t)$. This quickly converges to the EE optimal power within a couple of iterations for different values of $P(0)$ and $\hat{P}(0)$. The curve variances for the full search and fixed power algorithms solely come from temporally correlated fading, vanishing as time elapses.

Fig. 3 illustrates the maximized EE by the proposed power control increases along with $N$. In addition, the EE grows as $\lambda_b$ increases. This numerical evaluation is based on stationary fading, i.e. $\lim_{t\rightarrow \infty}|g(t)|$ following Rice$(|\mu|, \eta)$ according to Lemma 1.


\section{Conclusion}
In this work, we have characterized the EE of a UDN taking into account the spatio-temporal dynamics arising from BS locations, channel information, and interference patterns. Closed-form expressions of the MF measure and the utility function have been derived by invoking MF interference convergence as a function of  BS/user density ratio, fading channel, and reception ball size, i.e. LOS dependency of networks. Remarkably our result allows operators to optimize their energy-efficient networks with a large number of BSs in a tractable manner under MF constraints. On the basis of the proposed framework, incorporating user mobility is our work in progress \cite{MobilMFGSG:GC16}. Considering other spatio-temporal SG settings and/or spatial MFG could also be an interesting topic.

\section*{Ackhowledgement}\small
This work was supported by Institute for Information \& communications Technology Promotion (IITP) grant funded by the Korea government (MSIP) (No.B0126-16-1017, Spectrum Sensing and Future Radio Communication Platforms).
\normalsize

\section*{Appendix -- Proof of Theorem 1}
The following proves Theorem 1 for $N=1$ by sequentially verifying the aforementioned three conditions for MF interference (replacing  $\lambda_b$ in this proof with $N^{\frac{1}{\alpha}} \lambda_b $ straightforwardly leads to the desired result for $N>1$).

\noindent1. \textsf{Exchangeability} -- With the aid of de Finetti's theorem \cite{Durett:Book}, i.i.d. states are invariantly permutable. Showing the considered channel states $h_i$'s are i.i.d. thus suffices to prove the exchangeability. 

Firstly, BS coordinates $z_i$'s as well as fading vector $g_i$'s are mutually independent, and so are their $h_i$'s. Secondly, the BSs' common rule of user associations and homogeneous admissible power control along with the identical fading distribution lead to identically distributed $h_i$'s, ensuring exchangeability.
\vspace{5pt}

\noindent2. \textsf{MF measure existence} -- It requires to prove an empirical measure $M_t$ of $h_i$'s within $b(o,R)$ converges to $m_t$. The following specifies the conditions at which this convergence holds under a UDN.

\vspace{5pt}
\begin{enumerate}[2-1)]
\item \textsf{MF measure existence for $R < \infty$}: 

For $\E[|g_0|^2]<\infty$ and $\lambda_u\rightarrow \infty$, consider ${\lambda_b}/\lambda_u \rightarrow \infty$.
\vspace{-5pt}
\small\begin{align}
\hspace{-20pt}\underset{\begin{subarray}{c} {\lambda_b} /\lambda_u \rightarrow \infty \\ \lambda_u \rightarrow \infty\end{subarray}}{\lim} \hat{I}_{p_a\lambda_b}(t) &\nn\\
&\hspace{-70pt}\overset{(a)}{\sim} \hspace{-5pt}\underset{\begin{subarray}{c} {\lambda_b}/\lambda_u \rightarrow \infty \\ \lambda_u \rightarrow \infty\end{subarray}}{\lim} \frac{1 }{{\lambda_b}^{\frac{\alpha}{2}}} \[\sum_{i=1}^{| \Phi^o_R\({\lambda_u}\)|}  \hspace{-10pt}P_{i}\l\{t, h_{i}(t, \lambda_u)\r\}|h_{i0}(t, \lambda_u)|^2 -S_{\lambda_u}(t)\]\\
&\hspace{-70pt}=  \hspace{-5pt} \underset{\begin{subarray}{c} {\lambda_b}/\lambda_u \rightarrow \infty \\ \lambda_u \rightarrow \infty\end{subarray}}{\lim} \hspace{-5pt} \frac{|\Phi_R^o\({\lambda_u}\)|}{ {\lambda_b}^\frac{\alpha}{2} } \frac{1}{|\Phi_R^o\({\lambda_u}\)|} \sum_{i=1}^{|\Phi_R^o\({\lambda_u}\)|}\hspace{-10pt} P_i\l\{t, h_i(t, \lambda_u)\r\}|h_{i0}(t,\lambda_u)|^2  \label{Eq:PfThm1}
\end{align}\normalsize
where $(a)$ follows from $\lim_{\lambda_b/\lambda_u\rightarrow \infty} p_a = \lambda_u/\lambda_b$ by Taylor expansion, in conjunction with the mapping theorem \cite{HaenggiSG}. The last step comes from $S_{\lambda_u}(t)\leq P_{\max} |g_0|^2 <\infty$ $a.s$. The last inequality is derived by applying Markov's inequality to $|g_0|^2$ as follows. 
\begin{align}
\lim_{v\rightarrow \infty}\Pr\(|g_0|^2\geq v\) \leq \lim_{v\rightarrow \infty}\E[|g_0|^2]/v = 0
\end{align}

Define $M_t$ as 
\vspace{-5pt}\small\begin{align}
M_t &:= \frac{1}{|\Phi_R^o\({\lambda_u}\)|}\sum_{i=1}^{|\Phi^o_R\({\lambda_u}\)|}\delta_{h_i(t, \lambda_u)}.
\end{align}\normalsize

Note that $|\Phi^o_R\({\lambda_u}\)|$ is the number of active BSs within $b(o,R)$, following Poisson distribution with mean ${\lambda_u}\pi R^2$. It is therefore straightforward $|\Phi^o_R\({\lambda_u}\)|\rightarrow \infty$ $a.s.$ as $\lambda_u \rightarrow \infty$. In such a condition, the empirical measure $M_t$ converges to Borel probability measure $m_t$, i.e. $m_t= \lim_{\lambda_u \rightarrow \infty}M_t$ $a.s.$ according to Theorem 11.4.1 in \cite{Dudley:Book}. This proves the existence of $m_t$.

%

\vspace{5pt}
\item \textsf{MF measure existence for $R\rightarrow \infty$}:

The proof directly follows from (33) while relaxing the necessary condition by replacing $\lambda_u\rightarrow \infty$ with $\lambda_u>0$. Under this condition, $|\Phi_R^o(\lambda_u)|\rightarrow \infty$ as $R\rightarrow \infty$. This leads to $\lim_{R\rightarrow \infty} M_t = m_t$ $a.s.$

\end{enumerate}

\vspace{5pt}
\noindent 3. \textsf{Finite MF interference} -- In the following, the required BS density scaling laws under $R<\infty$ and $R\rightarrow \infty$ are provided so as to verify the convergence of MF interference in \eqref{Eq:PfThm1}.

\vspace{5pt}
\begin{enumerate}[3-1)]
\item \textsf{Finite MF interference for $R<\infty$}:

Consider $\lambda_b/{\lambda_u}^\frac{4}{\alpha} \rightarrow \infty$ for $\lambda_u \rightarrow \infty$ and $\E[|g_0|^2]<\infty$. Applying $m_t$, the terms after $|\Phi_R^o\({\lambda_u}\)| / \({\lambda_b}/\lambda_u\)^\frac{\alpha}{2} $ in \eqref{Eq:PfThm1} becomes


\vspace{-5pt}
\small\begin{align}
\frac{1}{|\Phi_R^o\({\lambda_u}\)|} \sum_{i=1}^{|\Phi_R^o\({\lambda_u}\)|} P_i\l\{t, X_i(t, \lambda_u)\r\}|h_{i0}(t,\lambda_u)|^2  &\nn\\
&\hspace{-200pt}\overset{(b)}{=}\E_\chi\[P\{t, x\}\]  \E_g\[|g(t)|^2\] \E_{\Phi_R^o\({\lambda_u}\)}\[l_{i0}(\lambda_u)\] \label{Eq:PfThm1-3}
\end{align}\normalsize
where $\E_\chi\[P\{t, x\}\] = \int_{x\in\chi} P\{t, x\} m_t(dx)$. The step $(b)$ comes from the mutual independence of $g_{i0}(t)$'s and $\ell_{i0}(\lambda_u)$'s. 

The first term $\E_\chi\[P\{t, x\}\]$ in \eqref{Eq:PfThm1-3} converges since $\E_\chi\[P\{t, x\}\] \leq P_{\max}   <\infty$ when considering the maximum power transmission while neglecting a generic BS's remaining energy state. The second term $\E_g\[| g(t) |^2\]$ converges by definition. By using Campbell's theorem \cite{HaenggiSG}, the last term is 

\vspace{-5pt}\small\begin{align}
\E_{\Phi_R^o\({\lambda_u}\)}\[l_{i0}(\lambda_u)\] &= {\lambda_u}\pi\(1 + \frac{1-R^{2-\alpha}}{\alpha-2}\).
\end{align}\normalsize

Applying this to \eqref{Eq:PfThm1-3} combined with $\lim_{\lambda_u\rightarrow \infty}|\Phi_R^o(\lambda_u)| = \E \[| \Phi_R^o(\lambda_u)|\]=\lambda_u \pi R^2$ makes \eqref{Eq:PfThm1} become

\vspace{-5pt}\small\begin{align}
\underbrace{\frac{(\lambda_u \pi R)^2}{{\lambda_b}^\frac{\alpha}{2}}}_{(c)} \(1 + \frac{1-R^{2-\alpha}}{\alpha-2}\)\E_\chi\[P\{t, x\}\]  \E_g\[|g(t)|^2\] .
\end{align}\normalsize


The term $(c)$ converges on $0$ since $\lambda_b/{\lambda_u}^\frac{4}{\alpha} \rightarrow \infty$, verifying $\hat{I}_{p_a\lambda_b}$ is $a.s.$ finite.

%


\vspace{5pt}
\item \textsf{Finite MF interference for $R\rightarrow\infty$}:

Consider $\lambda_b/(\lambda_u R)^\frac{4}{\alpha} \rightarrow \infty$ for $\lambda_u>0$ and $\E[|g_0|^2]<\infty$. 
Under these conditions, (38) holds and the term $(c)$ converges on $0$, yielding the desired result.

\end{enumerate}

Combining the results of subsections 1, 2, and 3 completes the proof of Theorem 1.  \endproof

\bibliographystyle{ieeetr}

\begin{thebibliography}{10}

\bibitem{Andrews5G:14}
J.~G. Andrews, S.~Buzzi, W.~Choi, S.~Hanly, A.~Lozano, A.~C.~K. Soong, and
  J.~C. Zhang, ``{What Will 5G Be?},'' {\em IEEE Journal on Selected Areas in
  Communications}, vol.~PP, no.~99, 2014.

\bibitem{FiveDisruptive:14}
F.~Boccardi, R.~W. Heath, Jr, A.~Lozano, T.~Marzetta, and P.~Popovski, ``{Five
  Disruptive Technology Directions for 5G},'' {\em IEEE Communications
  Magazine}, vol.~52, no.~2, pp.~74--80, 2014.

\bibitem{CLI:16}
L.~Su, C.~Yang, and C.-L. I, ``{Energy and Spectral Efficient Frequency Reuse
  of Ultra Dense Networks},'' {\em IEEE Transactions on Wireless
  Communications}, vol.~15, no.~8, pp.~5384--5398, 2016.

\bibitem{Youssef:16}
M.~Kamel, W.~Hamouda, and A.~Youssef, ``{Ultra-Dense Networks: A Survey},''
  {\em to appear in IEEE Communications \& Tutorials [Online]. Early access:
  http://ieeexplore.ieee.org/document/7476821}.

\bibitem{JHParkTWC:15}
J.~Park, S.-L. Kim, and J.~Zander, ``{Tractable Resource Management with Uplink
  Decoupled Millimeter-Wave Overlay in Ultra-Dense Cellular Networks},'' {\em
  IEEE Transactions on Wireless Communications}, vol.~15, no.~6,
  pp.~4362--4379, 2016.

\bibitem{Alexiou13}
A.~G. Gotsis and A.~Alexiou, ``{On Coordinating Ultra-Dense Wireless Access
  Networks: Optimization Modeling, Algorithms and Insights},'' {\em available
  at: http://arxiv.org/pdf/1312.1577v1.pdf}.

\bibitem{Meriaux:13}
F.~M{\'e}riaux, S.~Lasaulce, and H.~Tembine, ``{Stochastic Differential Games
  and Energy-Efficient Power Control},'' {\em Dynamic Games and Applications},
  vol.~3, no.~1, pp.~3--23, 2013.

\bibitem{Tembine:11}
H.~Tembine and M.~Huang, ``{Mean Field Difference Games: McKean-Vlasov
  Dynamics},'' {\em Proc. IEEE Conference on Decision and Control and European
  Control Conference (CDC-ECC), Orlando, FL, USA}, December 2011.

\bibitem{MBennisGC:15}
S.~Samarakoon, M.~Bennis, W.~Saad, M.~Debbah, and M.~Latva-aho,
  ``{Energy-Efficient Resource Management in Ultra-Dense Small Cell Networks: A
  Mean-Field Approach},'' {\em Proc. IEEE Global Communications Conference
  (GLOBECOM), San Diego, CA, USA}, December 2015.

\bibitem{StoyanBook:StochasticGeometry:1995}
D.~Stoyan, K.~W. S., and J.~Mecke, {\em {Stochastic Gemoetry and its
  Applications}}.
\newblock Wiley, 2nd~ed., 1995.

\bibitem{HeathWearable:15}
K.~Venugopal, M.~C. Valenti, and R.~W. Heath, Jr, ``{Interference in
  Finite-Sized Highly Dense Millimeter Wave Networks},'' {\em in Proc. ITA},
  Feb. 2015, pp. 175--180.

\bibitem{Honig:12}
M.~Agarwal, M.~L. Honig, and B.~Ata, ``{Adaptive Training for Correlated Fading
  Channels With Feedback},'' {\em IEEE Transactions on Information Theory},
  vol.~58, no.~8, pp.~5398--5417, 2012.

\bibitem{Heath:13}
T.~Bai, R.~Vaze, and R.~W. Heath, Jr, ``{Analysis of Blockage Effects on Urban
  Cellular Networks},'' {\em IEEE Transactions on Wireless Communications},
  vol.~13, no.~9, pp.~5070--5083, 2014.

\bibitem{Andrews:2011bg}
J.~G. Andrews, F.~Baccelli, and R.~K. Ganti, ``{A Tractable Approach to
  Coverage and Rate in Cellular Networks},'' {\em IEEE Transactions on
  Communications}, vol.~59, no.~11, pp.~3122--3134, 2011.

\bibitem{Yu2011}
S.~M. Yu and S.-L. Kim, ``{Downlink Capacity and Base Station Density in
  Cellular Networks},'' {\em Proc. IEEE WiOpt Workshop on Spatial Stochastic
  Models for Wireless Networks (SpaSWiN)}, May 2013.

\bibitem{SLeeKHuang12}
S.~Lee and K.~Huang, ``{Coverage and Economy of Cellular Networks with Many
  Base Stations},'' {\em IEEE Communications Letters}, vol.~16, no.~7,
  pp.~1038--1040, 2012.

\bibitem{HaenggiSG}
M.~Haenggi, {\em {Stochastic Geometry for Wireless Networks}}.
\newblock Cambridge University Press, 2013.

\bibitem{Kantorovich:48}
L.~V. Kantorovich, ``{Functional Analysis and Applied Mathematics},'' {\em
  Uspekhi Mat. Nauk}, vol.~3, no.~6(28), pp.~89--185, 1948.

\bibitem{MobilMFGSG:GC16}
J.~Park, S.~Jung, S.-L. Kim, M.~Bennis, and M.~Debbah, ``{User-Centric Mobility
  Management in Ultra-Dense Cellular Networks under Spatio-Temporal
  Dynamics},'' {\em to appear in Proc. IEEE Global Communications Conference
  (GLOBECOM) 2016, available at: http://arxiv.org/abs/1606.05673}.

\bibitem{Durett:Book}
R.~Durett, {\em {Probability: Theory and Examples}}.
\newblock Cambridge University Press, 4th~ed., 2010.

\bibitem{Dudley:Book}
R.~M. Dudley, {\em {Real Analysis and Probability}}.
\newblock Cambridge University Press, 2nd~ed., 2012.

\bibitem{Chernoff:Book}
M.~Mitzenmacher and E.~Upfal, {\em {Probability and Computing: Randomized
  Algorithms and Probabilistic Analysis}}.
\newblock Cambridge University Press, 2005.

\end{thebibliography}

\end{document}